\documentclass[]{interact}

\usepackage{epstopdf}% To incorporate .eps illustrations using PDFLaTeX, etc.
\usepackage{tabularx, booktabs}
\usepackage[caption=false]{subfig}% Support for small, `sub' figures and tables

\usepackage[numbers,sort&compress]{natbib}% Citation support using natbib.sty
\bibpunct[, ]{[}{]}{,}{n}{,}{,}% Citation support using natbib.sty
\theoremstyle{plain}% Theorem-like structures provided by amsthm.sty

\theoremstyle{definition}

\theoremstyle{remark}

\begin{document}

\articletype{ARTICLE TEMPLATE}% Specify the article type or omit as appropriate

\title{Unveiling Gamer Archetypes through Multi modal feature Correlations and Unsupervised Learning}

\author{
\name{Moona Kanwal\textsuperscript{a}\thanks{CONTACT A.~N. Author. Email: moona.kanwal@iqra.edu.pk}, Muhammad Sami Siddiqui\textsuperscript{b} and Syed Anael Ali\textsuperscript{c}}
\affil{\textsuperscript{a,b} FEST, Iqra University, Karachi, PK; \textsuperscript{c}Karachi Grammar School, PK}
}

\maketitle

\begin{abstract}
\begin{abstract}
	Profiling gamers provides critical insights for adaptive game design, behavioral understanding, and digital well-being. 
	This study proposes an integrated, data-driven framework that combines psychological measures, behavioral analytics, 
	and machine learning to reveal underlying gamer personas. A structured survey of 250 participants, including 113 active 
	gamers, captured multidimensional behavioral, motivational, and social data. The analysis pipeline integrated feature 
	engineering, association-network (knowledge-graph) analysis, and unsupervised clustering to extract meaningful patterns. 
	Correlation statistics (Cramér’s V, Tschuprow’s T, Theil’s U, and Spearman’s $\rho$) quantified feature associations, 
	and network centrality guided feature selection. Dimensionality-reduction techniques (PCA, SVD, t-SNE) were coupled with 
	clustering algorithms (K-Means, Agglomerative, Spectral, DBSCAN), evaluated using Silhouette, Calinski–Harabasz, and 
	Davies–Bouldin indices. The PCA + K-Means ($k = 4$) model achieved optimal cluster quality (Silhouette $\approx 0.4$), 
	identifying four archetypes: \textit{Immersive Social Story-Seekers}, \textit{Disciplined Optimizers}, 
	\textit{Strategic Systems Navigators}, and \textit{Competitive Team-Builders}.
	
	This research contributes a reproducible pipeline that links correlation-driven network insights with unsupervised learning. 
	The integration of behavioral correlation networks with clustering not only enhances classification accuracy but also offers 
	a holistic lens to connect gameplay motivations with psychological and wellness outcomes.
\end{abstract}

\end{abstract}


\section{Introduction}

Video gaming has evolved from a niche pastime into a global medium of entertainment and social interaction. While leisure activities have historically reflected cultural values and technological change, gaming has remained a constant, shaping identity and cultural expression from simple physical challenges to complex digital experiences. Today, video games play a central role in youth culture, offering immersive experiences that foster creativity, self-expression, and exploration of identity~\cite{huizinga1955homo} ~\cite{granic2014benefits}.

Research increasingly focuses on the psychological, social, and behavioral implications of gaming, including motivations, preferences, and well-being. A critical development in this field is gamer profiling. This profiling models player behavior to inform personalized game design, adaptive difficulty, and targeted interventions for both beneficial and potentially harmful outcomes such as addiction or aggression~\cite{ryan2006motivational, hamari2014player, vahlo2017digital}.

\begin{figure}[h]
	\centering
	\includegraphics[scale=0.25]{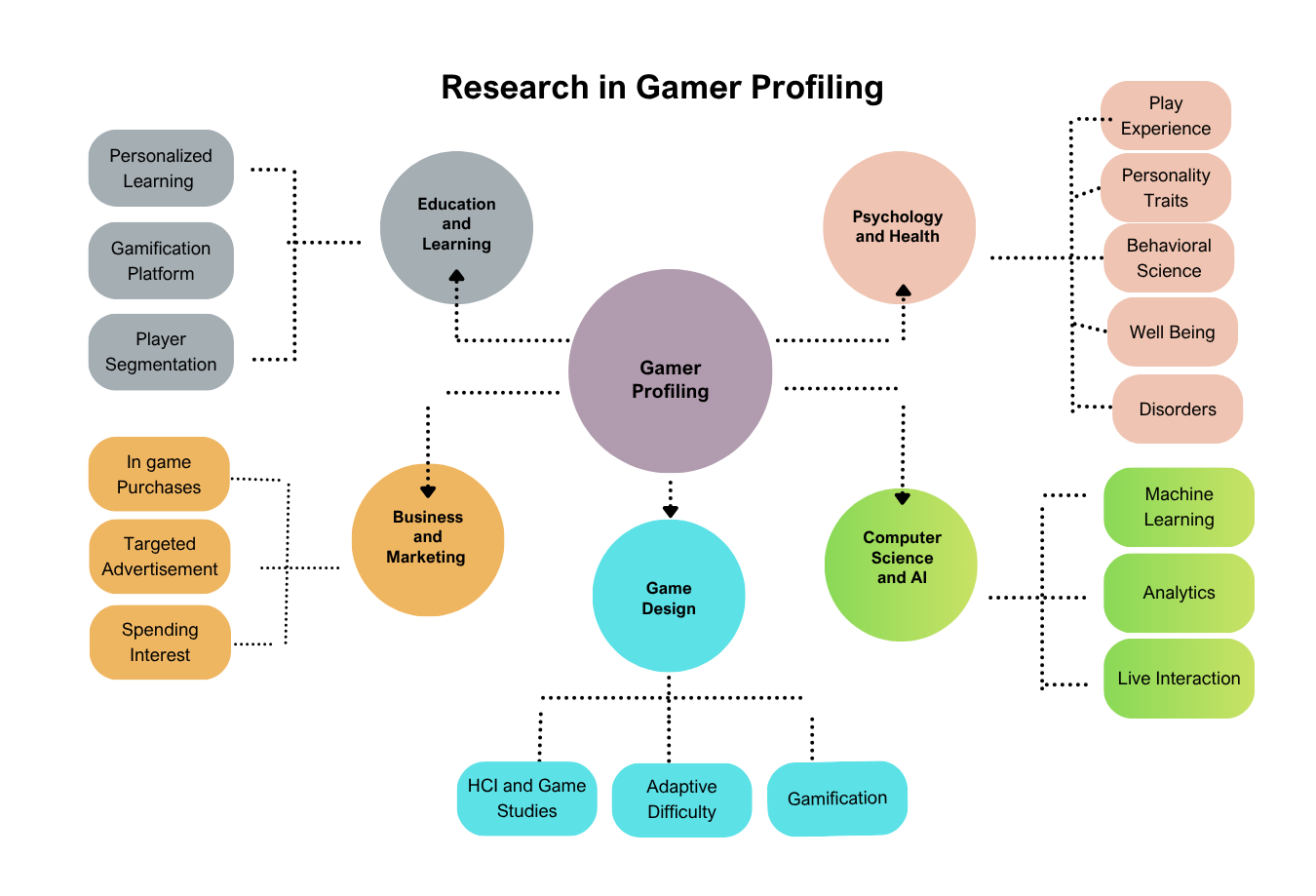}
	\caption{Landscape of research directions in gamer profiling.}
	\label{gamer_profiling}
\end{figure}

To clarify the broad landscape of gamer profiling research, Fig. ~\ref{gamer_profiling} presents a summary view of major sub-fields, including computer science, educational, psychological, business, and technological perspectives, demonstrating how this field draws upon and informs multiple domains.

Recent advances emphasize the use of telemetry and data-driven methods over self-reported measures, with machine learning enabling the analysis of large, complex datasets. Techniques such as unsupervised learning, natural language processing, and large language models provide new insight into player behavior~\cite{drachen2013comparison, krittanawong2021artificial}.

The objective of this research is to leverage advanced statistical methods and machine learning approaches to analyze the multimodal aspects of online gaming. Specifically, this research aims to develop an ML model to identify and identify different types of gamers based on theoretical frameworks spanning behavioral traits, psychological factors, sociological influences, personal characteristics, monetization strategies, self-regulation, engagement patterns, well-being, and gaming preferences.

By integrating statistical dependence measures, structural knowledge graph analysis, and robust clustering techniques, this work advances current understanding of gamer segmentation and motivation. The approach enables the discovery of distinct gamer personas and the behavioral dynamics underlying gaming engagement.

This research seeks to address the following research questions:
\begin{itemize}
	\item \textbf{RQ1:} How can multimodal data and theory-driven features inform the segmentation of online gamers?
	\item \textbf{RQ2:} What are the key behavioral, psychological, and social patterns that distinguish gamer types?
	\item \textbf{RQ3:} How can machine learning techniques be employed to model the relationships between engagement, self-regulation, and well-being in relation to gaming preferences across segments?
\end{itemize}
The paper is organized as follows: Section~2 reviews related literature, Section~3 outlines the methodology, and followed by Section~4 presents contribution, limitation future work and conclusion.

\section{Literature Review}
Understanding player profiling has become a central focus in game analytics research, evolving from clustering-based persona modeling to multifaceted applications in education, health, marketing, and artificial intelligence.

\subsection{Education \& Learning}

Recent literature on gamer profiling in educational games demonstrates the value of clustering and behavioral analytics for adaptive learning. Player-type models reveal that learners naturally group by distinct motivational and behavioral tendencies, supporting the tailoring of game mechanics to maximize engagement and educational impact~\cite{brandl2024student}. Studies employing analytics, such as clustering on in-game activity, sequence analysis, and dashboard-based visualization, can identify patterns ranging from hint reliance and independent exploration to varied retry and error trajectories~\cite{dai2023exploring, wang2024discovering, calvo2025learning}. These profiles empower educators not only to personalize interventions and adapt difficulty, but also to iteratively refine digital game design itself~\cite{banihashem2024learning} ~\cite{calvo2025learning}. Overall, profiling approaches yield actionable insights and facilitate more effective, responsive, and engaging game-based learning experiences.

\subsection{Business \& Marketing}
Commercial analytics deploy profiling for purchase prediction and marketing. Cengiz et al.~\cite{cengiz2025competitive} link competitive gaming traits with impulse buying using surveys and transaction records, and Santos et al. ~\cite{Santos2024} conducted a comprehensive bibliometric and TCCM  analysis on 114 articles from the Scopus database, providing an integrated overview of gamification in marketing.
. Industry workshops (~\cite{compsac2025workshops}) showcase advanced AI and analytics driving adaptive content and personalization at scale.

\subsection{Gamer Motivations, Typologies, and Health}

\paragraph{Gamer Motivations and Typologies}

Recent research on gamer motivations and typologies has combined psychometric questionnaires, empirical surveys, and clustering approaches to deepen understanding of player psychology. Foundational work by Bartle~\cite{bartle1996mud} used qualitative analysis to establish four archetypes: Achiever, Socializer, Explorer, and Killer, setting the stage for later empirical studies. Yee~\cite{yee2006motivations} applied large-scale survey factor modeling to identify achievement, social, and immersion as dominant online gaming motivations. Demetrovics et al.~\cite{demetrovics2011you} developed the Motives for Online Gaming Questionnaire (MOGQ), which quantifies psychological drivers such as competition, achievement, social interaction, and immersion. Kiraly et al.~\cite{kiraly2022comprehensive} further refined motivation assessment with the Gaming Motivation Inventory (GMI), which distinguishes between adaptive and maladaptive motives tied to wellness outcomes.

\paragraph{Personality and Neurobiological Perspectives}

Integrating personality frameworks and neurobiology, BrainHex by Nacke et al.~\cite{nacke2014brainhex} categorized players into seven archetypes related to neurobiological responses. Vera Cruz et al.~\cite{veracruz2023bigfive} mapped Big Five personality traits onto gaming behaviors, illustrating how traits influence play styles. Kahila et al.~\cite{kahila2023metagamers} complemented these insights through typological analysis, identifying “metagamer” profiles defined by creative and strategic engagement beyond conventional gameplay. George and Ranjith~\cite{george2024emotional} demonstrated through psychological surveys and network analysis that intrinsic motivation and emotional bonds drive engagement and well-being among casual and esports gamers.

\paragraph{Health and Well-being Implications}

Gamer profiles also closely relate to health outcomes such as stress, resilience, and risk of problematic gaming. Castro and Neto~\cite{castro2025profiling} identified four profiles relevant for mental health interventions via psychometric surveys and behavioral clustering. Aonso-Diego et al.~\cite{aonsodiego2024depression} used latent profile analysis to associate depression and anxiety with gaming habits, while Canale et al.~\cite{canale2025problem} revealed longitudinal links between adolescent gaming, well-being, and problematic behaviors. Personality traits like openness and conscientiousness predict preferences for healthy gaming, whereas neuroticism and impulsivity align with elevated risks of pathological play~\cite{potard2022video, dehesselle2021association}. Di Cesare et al.~\cite{diCesare2024sgdy} and Pitroso et al.~\cite{pitroso2024gamingcommunities} explored how gender, subcultures, and microaggressions within gaming communities shape psychological and social outcomes. Johannes et al.~\cite{johannes2021video} found that objective telemetry data often outperform self-reports in predicting well-being, showing positive correlations between gaming time and subjective life satisfaction.

\paragraph{Integrative Clustering Approaches}

Recent profiling research has employed clustering methods integrating motivational, affective, and demographic data to derive comprehensive gamer typologies. Kim et al.~\cite{kim2023latent} used latent profile analysis with high entropy to classify internet gamers along psychological and behavioral dimensions, revealing unique risk and motivation patterns. This integrative trend unifies motivational theories, personality insights, and health outcomes, providing a holistic framework for understanding gamer behavior and its impacts.

\subsection{Computer Science \& AI}
Early studies such as Salminen et al.~\cite{salminen2020designing} use k-means clustering on behavioral and demographic game data to prototype player personas, highlighting demographic and motivational differences. De Simone et al.~\cite{desimone2021design} develop collaborative recommender systems using game preference data and latent feature extraction. It demonstrates the impact of personalization on engagement and satisfaction. The researchers reviewed various artificial intelligence models, including supervised methods such as SVM, random forests, and neural networks~\cite{hashemi2022artificial}, as well as unsupervised approaches like clustering algorithms~\cite{drachen2013comparison}, applied to behavioral data. They demonstrated that both the choice of method and the incorporation of behavioral features significantly enhance model performance and robustness. Bowditch et al.~\cite{bowditch2025gamertypology} apply typology frameworks to large datasets, revealing important links between player subtypes, well-being, and online behavior.

\subsection{Game Design}
Modern profiling research further informs dynamic game design and engagement modeling. Acharya et al.~\cite{acharya2021predicting} develop engagement prediction models in online games using XGBoost on telemetry logs, showing that boosted models outperform traditional approaches. Gosztonyi~\cite{gosztonyi2022profiling} undertakes gamer profiling in a semi-peripheral EU nation via survey analytics and clustering, identifying unique subcultural dynamics and motivational patterns. Setiawan et al.~\cite{setiawan2021player} use a Gaussian Naive Bayes classifier on game logs and survey responses, achieving an accuracy of approximately 84.3\% to effectively capture engagement patterns in noisy or imbalanced data.

\vspace{0.5em}

Together, these interdisciplinary findings establish that gamer profiling not only advances the science of play and digital culture but also underpins next-generation solutions in education, business, health, and interactive technology.
	\section{Methodology}
\begin{figure}
	\centering
	\includegraphics[width=0.7\linewidth]{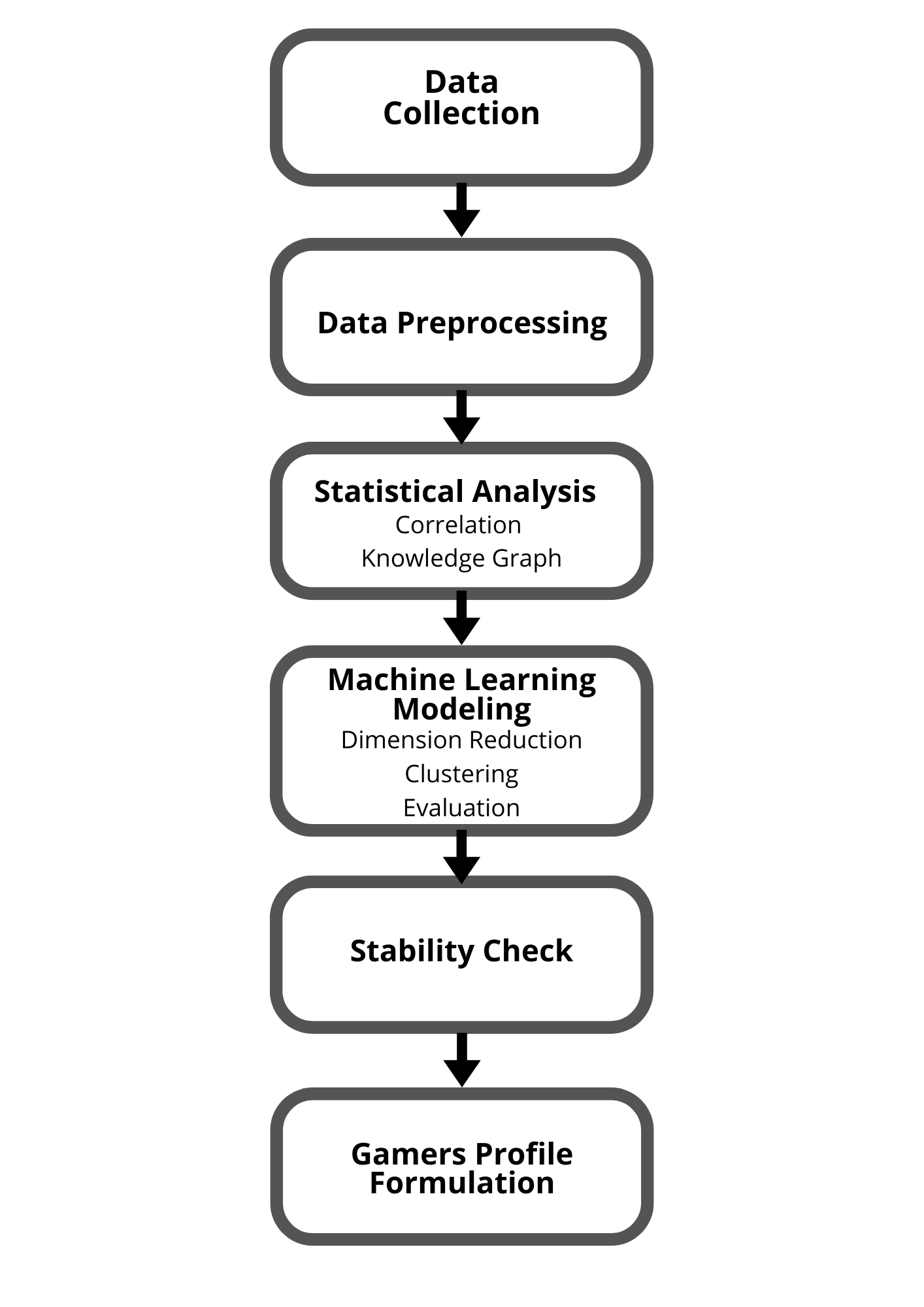}
	\caption{Flowchart of research}
	\label{fig2}
\end{figure}

This study implements a structured, multi-stage analytic pipeline to uncover and profile the diverse psychological, behavioral, and motivational characteristics of gamers. As illustrated in Fig.~\ref{fig2}, the methodology encompasses survey design and data collection, rigorous data cleaning and encoding, statistical correlation analysis, knowledge graph construction, unsupervised machine learning, and robust cluster profiling. Each stage ensures reproducibility and supports the discovery of meaningful gamer personas and behavioral patterns.

\subsection{Survey Design and Data Collection}
The survey is designed to profile gamer by assessing their psychological, behavioral, social, motivational, along with references and habits related to gaming.

It also collect demographic data (age, gender, academic discipline etc.) to provide a comprehensive player profile.

Questions includes Likert scales (five- and seven-point) for mood and attitudinal measures, categorical and multiple-choice items for nominal data like platform and genre preferences, and frequency-based interval questions to capture quantitative behavior.

A stratified random sampling approach recruited participants from university cohorts, online gaming communities, and gaming events to ensure a diverse and representative sample of casual and engaged gamers.
Participation is voluntary and anonymous, with informed consent and data privacy rigorously maintained following ethical guidelines. 
No personally identifiable information is collect. Data is gathered digitally via standardized platforms (e.g., Google Forms), ensuring consistent formatting for easy preprocessing and analysis.

\subsection{Data Pre-processing}
\begin{figure}[h]
	\centering
	\includegraphics[width=0.6\linewidth]{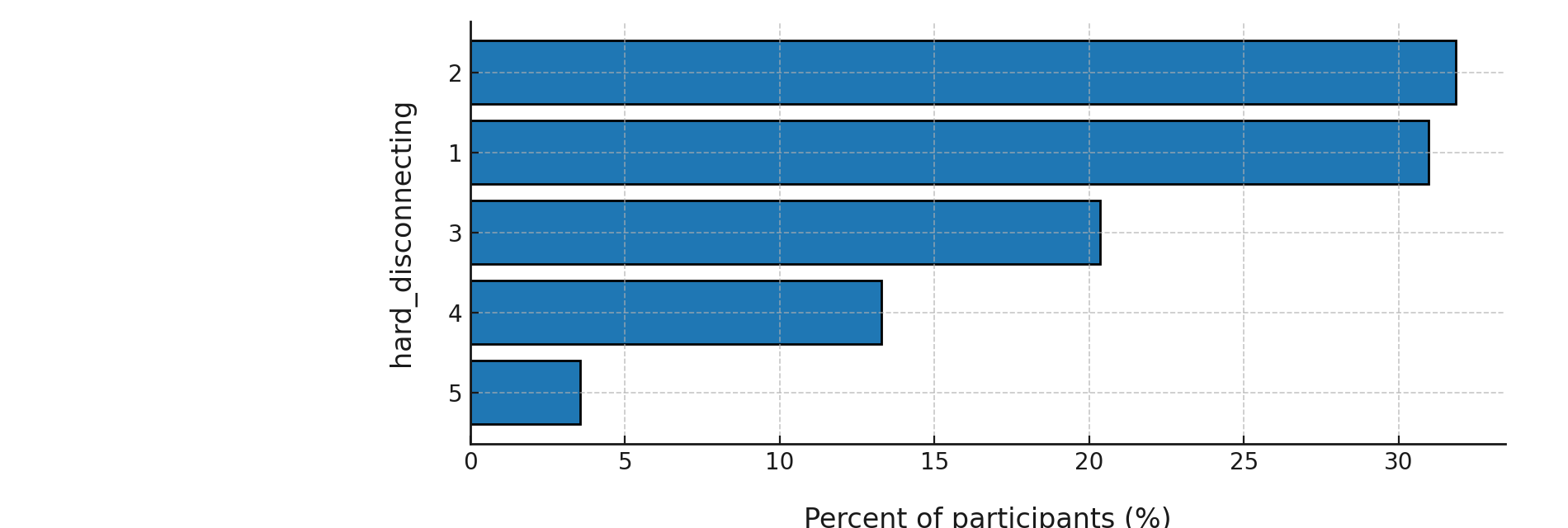}
	\includegraphics[width=0.4\linewidth]{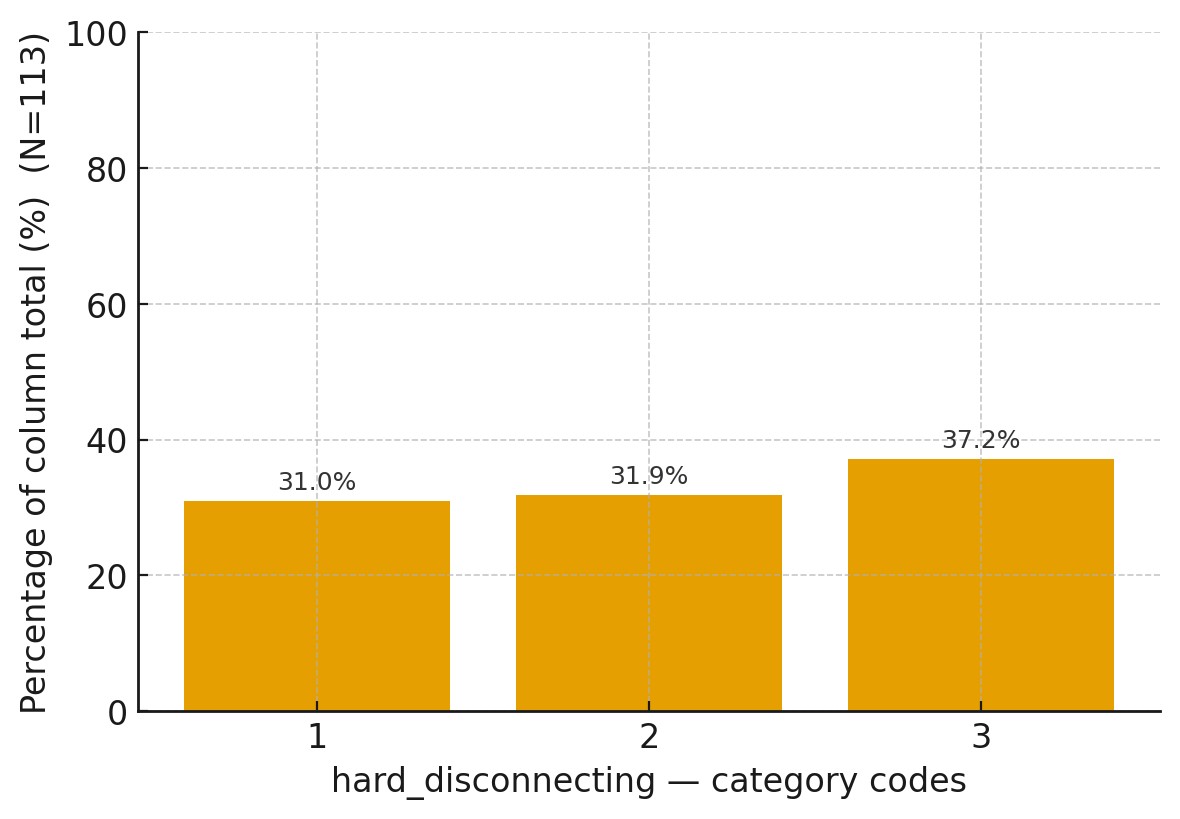}
	\caption{(a) Positive skewness in \texttt{hard\_disconnecting} feature before merging; (b) after merging categories.}
	\label{fig:skewness}
\end{figure}

A multi-stage data pre-processing workflow is employed to prepare the dataset.
Of 236 study participants, 113 are retained as gamers based on self-identification and reported play durations.
Multi-select responses are normalized by lowercasing text, trimming open-ended entries, removing within-row duplicates, and splitting on common delimiters (commas, slashes ``/'', ampersands ``\&'', plus signs ``+'', and ``and'')
Text aliases are harmonized prior to encoding, and noisy tokens are corrected or removed to avoid invalid features (e.g., slang merged into ``none'' and near-synonyms like ``chewing items'' mapped to ``eating'').
For example Fig.
~\ref{fig:skewness}(a) shows positive skewness in \textit{$hard-{disconnecting}$}, originally with five levels. Categories are merged into three levels (High, Moderate, Low Agree) as shown in Fig. ~\ref{fig:skewness}(b).

Noisy, redundant, or low-variance features were removed, including some demographic features such as occupation and gender. All preprocessing transformations were explicitly defined, reproducible, and applied uniformly across the dataset.

The dataset contained nominal and ordinal features. Nominal features are one-hot encoded, and Likert-scale items are label-encoded, using minimal transformations to preserve semantic integrity.
To reduce sparsity and enhance interpretability, granular one-hot columns are consolidated (e.g., hobbies merged into seven features, gaming habits into three features).

Compact numeric summaries are derived to capture repertoire breadth and effective state.
Gaming genre richness for participant $i$ is defined as the total number of distinct game genres selected by that individual, as given by $R_{\text{genre}}(i)$ in equation~\eqref{eq:genre}.

\begin{equation}
	R_{\text{genre}}(i) = \sum_{j=1}^J O_{ij},
	\label{eq:genre}
\end{equation}
where \(O_{ij} \in \{0,1\}\) indicates selection of genre \(j\). 

\textit{Family richness}$	R_{\text{family}}$ is defined as in ~\ref{eq:family}:
\begin{equation}
	R_{\text{family}}(i) = \sum_{k=1}^K G_{ik}, 
	\quad
	G_{ik} =
	\begin{cases}
		1 & \text{if } \sum_{j=1}^J O_{ij} M_{jk} \ge 1, \\
		0 & \text{otherwise.}
	\end{cases}
	\label{eq:family}
\end{equation}

where \(M_{jk} \in \{0,1\}\) maps genre \(j\) to family \(k\) 

Similarly, mood during gaming (\textit{$mood-{during}$}) is encoded into valence and arousal following affective measurement conventions ~\cite{potard2022video}. Valence: $V^+ = $ \{Excitement, Happy, Calmness, Contented\}, $V^0 =  $\{Neutral\}, $V^- = $ \{Anxiety, Depressed, Anger, Guilty\} as in ~\ref{eq:valence}. Arousal: $A^+ = $ \{Excitement, Happy, Anger, Anxiety, Guilty\}, $A^0 = $\{Neutral\}, $A^-= $ \{Calmness, Contented, Depressed\} as in ~\ref{eq:arousal}:
\begin{equation}
	\text{valence}(i) =
	\begin{cases}
		+1 & m_i \in V^+, \\
		0  & m_i \in V^0, \\
		-1 & m_i \in V^-.
	\end{cases}
	\label{eq:valence}
\end{equation}

\begin{equation}
	\text{arousal}(i) =
	\begin{cases}
		+1 & m_i \in A^+, \\
		0  & m_i \in A^0, \\
		-1 & m_i \in A^-.
	\end{cases}
	\label{eq:arousal}
\end{equation}

Noisy, redundant, or low-variance features (e.g., occupation, gender) are removed. All preprocessing steps are fully documented, reproducible, and applied consistently.

\subsection{Statistical Analysis}
A multi-method correlation analysis is conducted to examine the relationships among gamer-profiling variables and to extract significant features. 
The goal is to map how behavioral, psychological, and motivational factors interact across a mixed dataset, informing subsequent feature selection and persona segmentation.

\begin{figure}[hbt!]
	\centering
	\includegraphics[width=0.45\textwidth]{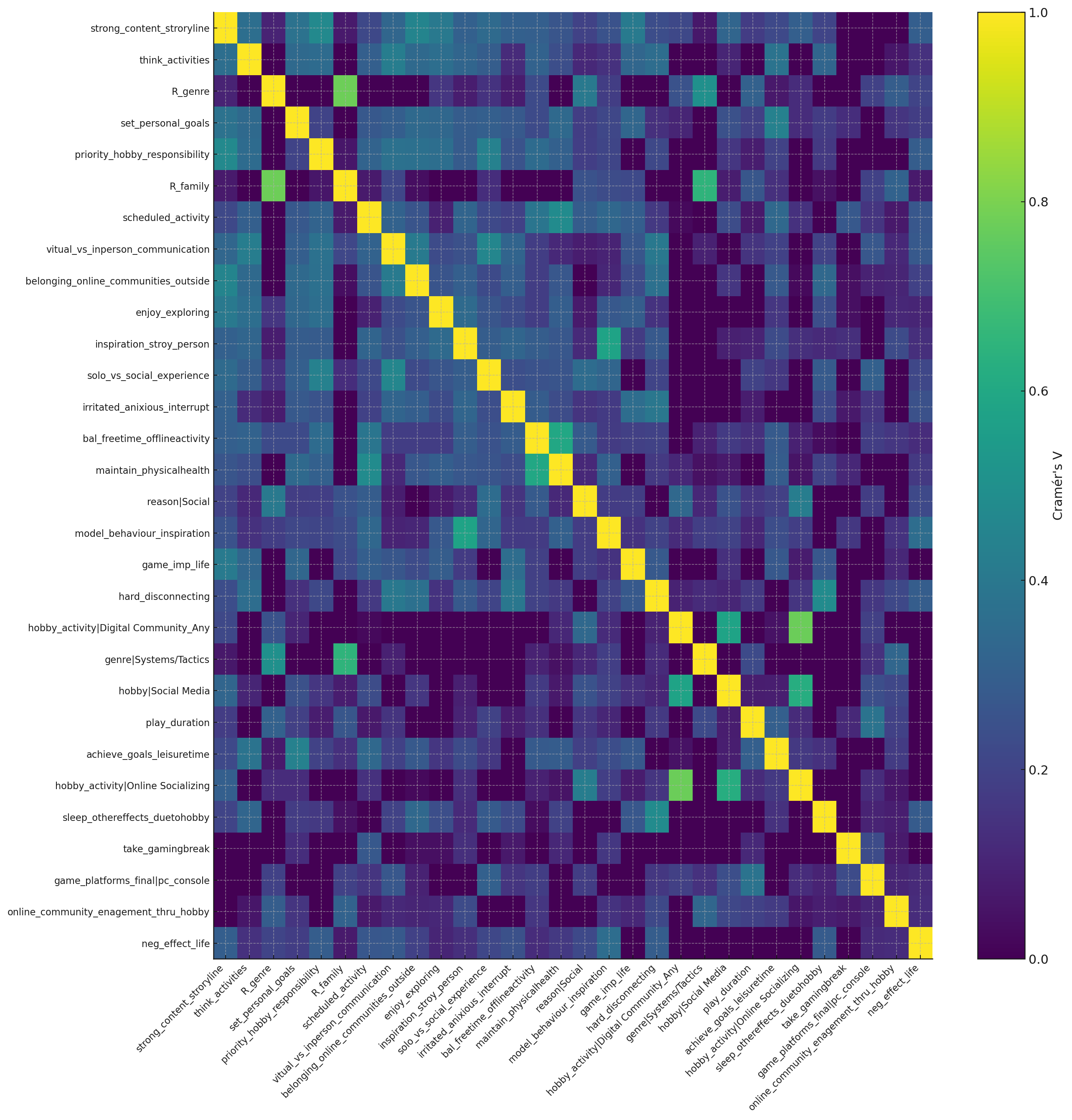}
	\includegraphics[width=0.45\textwidth]{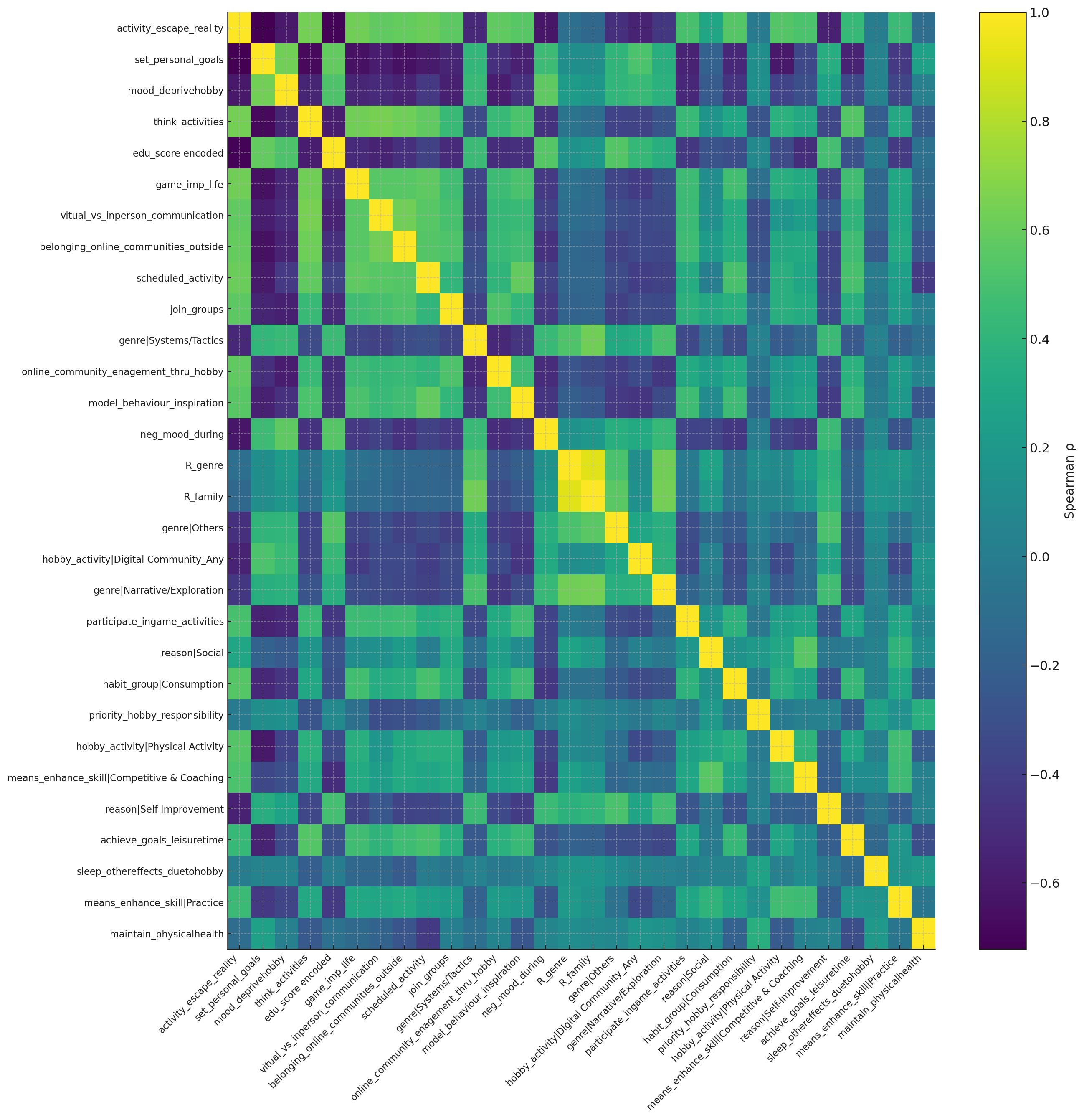}
	\caption{Correlation heatmaps of significant associations: (Up) Cramér's V; (Down) Spearman's $\rho$.}
	\label{fig:heatmaps}
\end{figure}
\begin{figure}[hbt!]
	\centering
	\includegraphics[width=0.45\textwidth]{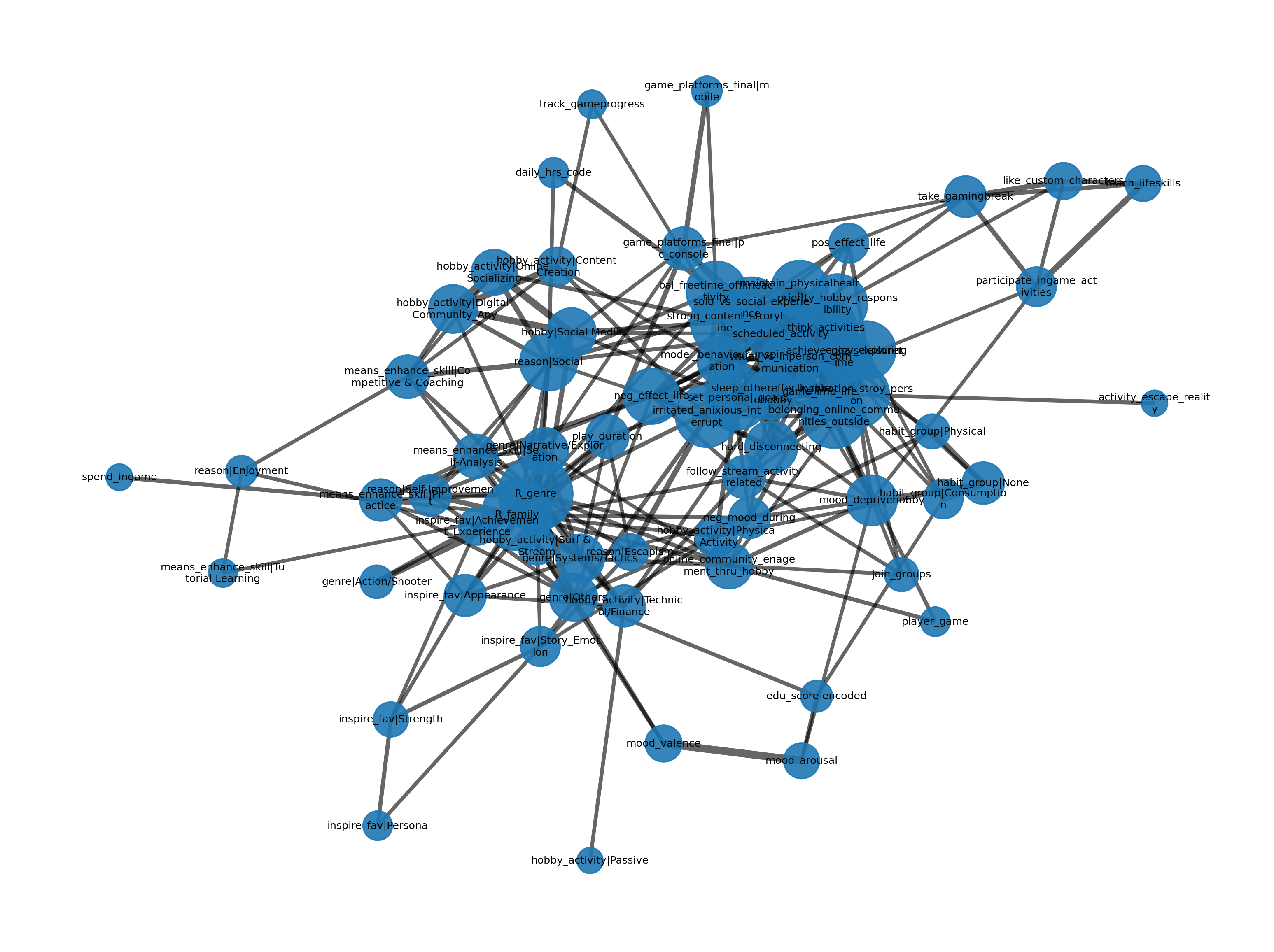}
	\includegraphics[width=0.45\textwidth]{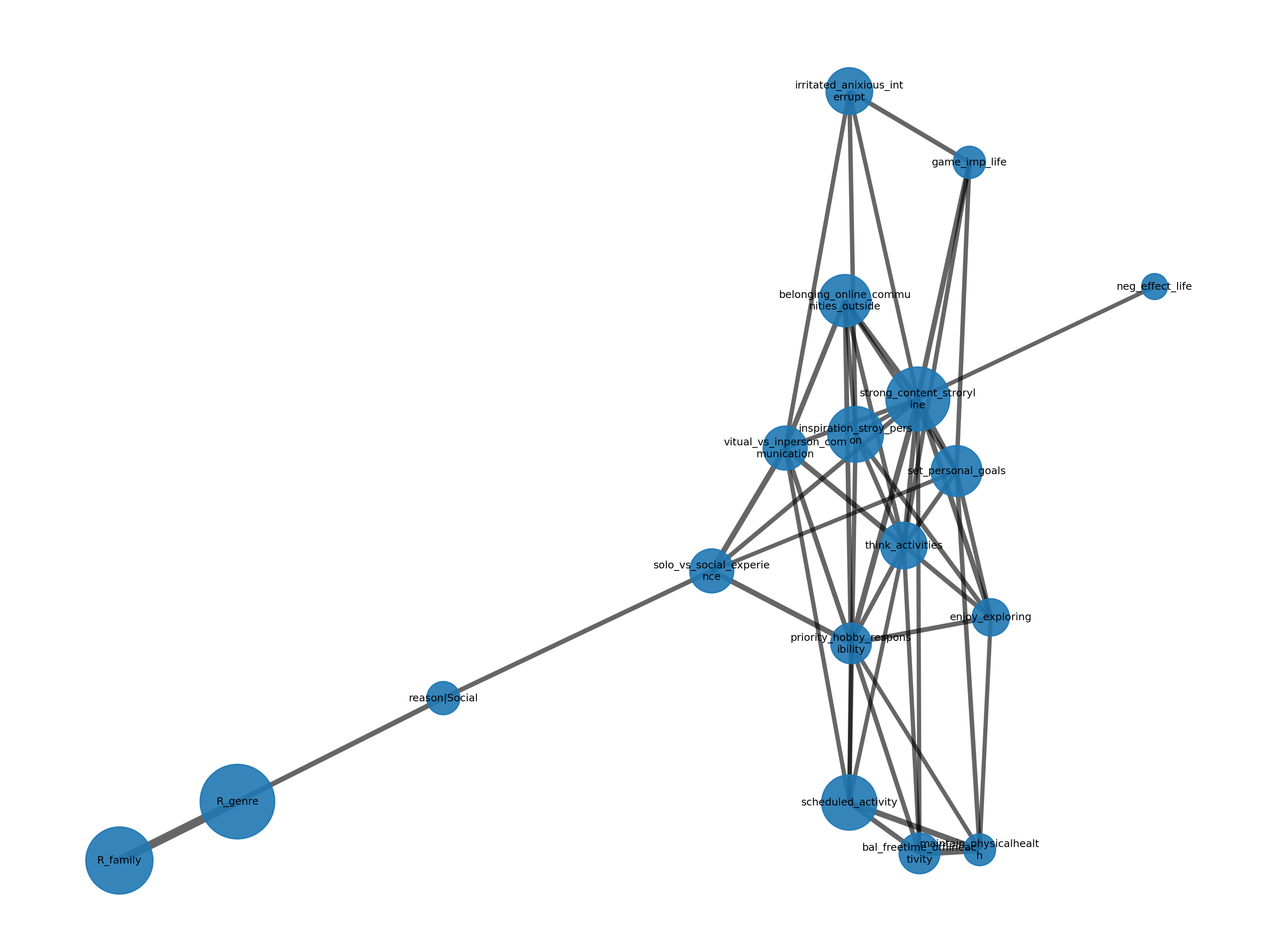}
	\caption{Knowledge graph (edges scaled by association weight); repertoire breadth and affect act as major hubs/bridges.}
	\label{fig:knowledgegraph}
\end{figure}

\subsubsection{Correlation}
Correlation analysis quantifies the strength and direction of relationships between variables. This study applies bias-corrected Cramér's V~\cite{agresti2012categorical} and Tschuprow's T~\cite{tschuprow1939theorie} for categorical associations, Theil's U~\cite{theil1972statistical} for directional predictability, Spearman's rank correlation coefficient ($\rho$)~\cite{spearman1904proof}  for monotonic relationships, and the chi-square ($\chi^2$) ~\cite{pearson1900x2} test for omnibus associations, expressed as $-\log_{10} p$-values.
%~\cite{johannes2021video}.
Numeric features—including genre family richness ($R_{\text{family}}$), genre richness ($R_{\text{genre}}$), mood arousal, and mood valence—are treated as continuous, enabling robust analysis alongside categorical items.

Visual correlation heatmaps (Figure~\ref{fig:heatmaps}) summarize the strongest associations identified for both categorical and ordinal/interval data.

Table~\ref{tab:thresholds} outlines the threshold values employed to filter significant relationships, in accordance with established criteria~\cite{johannes2021video}.

\subsubsection{Knowledge Graph and Network-Based Insights:}

To extend statistical analysis, a \textbf{knowledge graph framework}~\cite{hogan2021knowledge} is constructed:
\begin{itemize}
	\item \textbf{Nodes} represented survey features, and \textbf{edges} captured bias-corrected Cramér’s V associations.
	\item Edge filtering retained values $\geq 0.22$ (75th percentile).
	\item Strong-core subgraphs highlighted the most influential hubs with edges $\geq 0.30$.
	\item \textbf{Metrics}: degree, weighted degree (signal strength), and betweenness centrality (bridges).
	\item \textbf{Community detection} grouped features into interpretable clusters, such as narrative affinity, social engagement, and regulation/balance.
\end{itemize}

This knowledge graph facilitated the identification of central hubs, bridging nodes, and community structures within the gamer-profiling variables~\cite{fortunato2016community}.Fig.~\ref{fig:knowledgegraph} illustrates the resulting knowledge graph, emphasizing breadth of connections and affective variables as key central nodes and connectors within the network.

These observations provides a comprehensive understanding of intervariable dependencies, supporting subsequent feature selection and clustering analyses.

\begin{table}[h]
	\centering
	\scriptsize
	\renewcommand{\arraystretch}{3}
	\caption{Threshold values for correlation techniques used in association analysis}
	\label{tab:thresholds}
	\begin{tabular}{|p{1.7cm}|p{1cm}|p{4.8cm}|}
		\hline
		\textbf{Technique} & \textbf{Threshold} & \textbf{Rationale} \\
		\hline
		Chi-square $p$-value & 0.05 & False discovery rate control using BH procedure \\
		Spearman's $\rho$ & $\geq 0.50$ & Moderate or stronger monotonic association \\
		Cramér's V & $\geq 0.30$ & Moderate+ association for nominal data \\
		Tschuprow's T & $\geq 0.30$ & Comparable to Cramér's V for nominal variables \\
		Theil's U & $\geq 0.20$ & $\geq 20\%$ proportional reduction in uncertainty \\
		\hline
	\end{tabular}
\end{table}
\subsection{Machine Learning Approach}
Unsupervised machine learning pipeline is designed to identify latent behavioral patterns within mixed-type online gamer data. 
The methodological framework comprised four sequential stages: (i) dimensionality reduction to mitigate data complexity, (ii) implementation of clustering algorithms to partition the data into distinct groups, and (iii) internal validation to assess cluster (iv) stability and coherence of the resulting clusters

\subsubsection{Dimension Reduction}
Three complementary techniques, namely Principal Component Analysis (PCA), t-distributed Stochastic Neighbor Embedding (t-SNE), and Singular Value Decomposition (SVD), is employed to project the pruned feature set. Dimension reduction is also used to facilitate the clustering by providing a more representative description of the features.
PCA and SVD cover the linear, interpretable, reproducible side needed for clustering and reporting; t-SNE adds a non-linear lens for visualization, improving qualitative confidence without compromising metric validity.

There are other techniques like kernel PCA, MCA , FAMD etc but are not suitable due to higher risk of overfitting, reduce reproducibility at small data size or harder to interpret.

\paragraph{Principal Component Analysis (PCA)} For centered data matrix $\mathbf{X} \in \mathbb{R}^{n \times p}$ with covariance $\mathbf{S} = \frac{1}{n-1}\mathbf{X}^T\mathbf{X}$, PCA finds transformation $\mathbf{W} \in \mathbb{R}^{p \times d}$ maximizing retained variance~\cite{drachen2013comparison}. PCA finds top $d$ eigenvectors of $\mathbf{S}$ for projection. PCA(2D) is used mainly for visualization.

\paragraph{Singular Value Decomposition (SVD)} Decomposes $\mathbf{X}$ as $ \mathbf{X} = \mathbf{U} \boldsymbol{\Sigma} \mathbf{V}^T$ and gives rank-$r$ approximation $\mathbf{X}_r = \mathbf{U}_r \boldsymbol{\Sigma}_r \mathbf{V}_r^T$ which minimizes the Frobenius error: 
$\| \mathbf{X} - \mathbf{X}_r \|_F $~\cite{kim2023latent}. Up to 30 SVD components are used for high-dimensional encodings.

\paragraph{t-Distributed Stochastic Neighbor Embedding (t-SNE)} Constructs neighbor probabilities $p_{ij}$ in high dimension and $q_{ij}$ in low dimension, minimizing Kullback--Leibler divergence~\cite{kim2023latent}:
\begin{equation}
	C = \sum_i \sum_j p_{ij} 
	\log \left( \frac{p_{ij}}{q_{ij}} \right)
	\label{eq:kl_divergence}
\end{equation}
t-SNE(2D) helps reveal nonlinear structure missed by linear methods.

\subsubsection{Clustering Algorithms}
Multiple algorithms are applied on the reduced dimension in different combination:

\textbf{K-Means} is one of the most widely used unsupervised learning algorithms for partitioning data into $K$ clusters.  Mathematically, the optimization problem is defined as:
\begin{equation}
	\min \sum_{k=1}^{K} \sum_{\mathbf{x}_i \in C_k} 
	\left\| \mathbf{x}_i - \boldsymbol{\mu}_k \right\|^2
	\label{eq:kmeans}
\end{equation}

where clusters $\{C_k\}$ have centroids $\{\boldsymbol{\mu}_k\}$~\cite{salminen2020designing}.

\textbf{Agglomerative (Ward linkage)} method merges clusters to minimize within-cluster variance~\cite{drachen2013comparison}:
\begin{equation}
	\Delta(C_a, C_b) = 
	\frac{n_a n_b}{n_a + n_b} 
	\left\| \boldsymbol{\mu}_a - \boldsymbol{\mu}_b \right\|^2
	\label{eq:ward}
\end{equation}

\textbf{Spectral Clustering} constructs similarity matrix $\mathbf{W}$ and degree matrix $\mathbf{D}$, then the normalized Laplacian~\cite{kim2023latent}:
\begin{equation}
	\mathbf{L}_{\text{sym}} = 
	\mathbf{I} - 
	\mathbf{D}^{-1/2} \mathbf{W} \mathbf{D}^{-1/2}
	\label{eq:laplacian_sym}
\end{equation}

First $K$ eigenvectors are clustered using K-Means.

\textbf{DBSCAN (Density-Based Spatial Clustering)} is a density-based clustering algorithm that groups together points closely packed under a given metric $(X, d)$ with parameters $\varepsilon > 0$ and ${min\_samples} \geq 1$~\cite{setiawan2021player}. A point $p$ is a core point if its $\varepsilon$-neighborhood $N_\varepsilon(p) = \{ q \in X \mid d(p, q) \leq \varepsilon \}$ contains at least ${min\_samples}$ points. Clusters are formed by density-connected points, while points that do not belong to any cluster are labeled as noise ($-1$).

Four major clustering families—centroid-based, connectivity or variance-based, density-based, and graph/spectral methods—are employed to avoid overfitting conclusions to a single geometric assumption, while ensuring compatibility with the chosen embedding.

\subsubsection{Validation Metrics}
To evaluate model compactness and separation, we use three standard metrics.

\textbf{Silhouette Coefficient}~\cite{rousseeuw1987silhouettes} measures how well each point fits within its cluster compared to other clusters.
\begin{equation}
	s(i) = \frac{b(i) - a(i)}{\max \{ a(i), b(i) \}}, 
	\quad s(i) \in [-1,1]
	\label{eq:silhouette}
\end{equation}

where $a(i)$ is mean intra-cluster distance, $b(i)$ is lowest mean distance to other clusters.

\textbf{Calinski--Harabasz (CH) Index} ~\cite{calinski1974dendrite} evaluates clustering quality based on the ratio of between-cluster to within-cluster dispersion.
\begin{equation}
	\text{CH} = \frac{\mathrm{Tr}(\mathbf{B}_k)/(k-1)}{\mathrm{Tr}(\mathbf{W}_k)/(n-k)}
	\label{eq:CH}
\end{equation}

where $\mathbf{B}_k$ and $\mathbf{W}_k$ are between/within-cluster dispersion matrices.

\textbf{Davies--Bouldin (DB) Index}~\cite{davies1979cluster} quantifies average similarity between clusters, with lower values indicating better separation.
\begin{equation}
	\text{DB} = \frac{1}{k} \sum_{i=1}^k 
	\max_{j \neq i} \frac{S_i + S_j}{M_{ij}}
	\label{eq:DB}
\end{equation}

where $S_i$ is average intra-cluster distance for $i$, $M_{ij}$ is centroid distance between clusters $i$ and $j$.

\subsubsection{Stability}
Cluster stability is assessed using bootstrapping in combination with the \textit{Adjusted Rand Index (ARI)}~\cite{brouwer2009extending}.
%~\cite{setiawan2021player, hubert1985comparing}. 
The ARI provides a measure of agreement between different clustering solutions, where values close to 1 indicate perfect consistency, values near 0 suggest random agreement, and negative values reflect disagreement beyond chance. Additionally, the \textit{Jaccard Index}~\cite{tang2021evaluating} is applied to evaluate the robustness of the k-means algorithm in consistently identifying clusters across the full dataset, independent of initialization effects.
%~\cite{hastie2009elements}.
To further validate cluster stability, the cluster labels are appended to the dataset and used as target classes in supervised learning using logistic regression~\cite{hosmer2013applied}.
\section{Results and Discussion}

This section integrates statistical findings (the correlation structure and a knowledge graph), and the clustering pipeline derive gamer personas. The results are contextualized with reference to existing models and discuss implications for behavior and design.

\subsection{Sample Profile and Gameplay Characteristics}

The analytic sample consists of 113 self-identified gamers, filtered from approximately 250 respondents. The participant is a male student aged 16–23 from Pakistan. Median daily gaming time ranges from 2 to 3 hours, with some participants engaging up to 6 hours per day on average. Session lengths typically approximate 2 hours as in Fig.~\ref{fig:demographics}. Console and PC platforms are linked to longer sessions, while mobile gaming is popular but less predictive of genre preferences. Participants favor action and exploration genres, with a notable subset attracted to system and tactics gameplay (Table~\ref{tab:participant_chars}, Figure~\ref{fig:demographics})~\cite{veracruz2023bigfive}.

\begin{table}[h]
	\centering
	\caption{Participant Demographics and Gameplay Characteristics}
	\begin{tabular}{|l|l|}
		\hline
		Characteristic & Description \\
		\hline
		Gender & Predominantly male \\
		Age Range & 16–23 years \\
		Daily Gaming & Median 2–3 hours, upper tail ~6 hours \\
		Session Length & Approx. 2 hours \\
		Platform Usage & Console/PC tied to longer sessions \\
		Popular Genres & Action, Exploration, Systems/Tactics \\
		\hline
	\end{tabular}
	\label{tab:participant_chars}
\end{table}

\begin{figure}[h]
	\centering
	\includegraphics[width=0.48\textwidth]{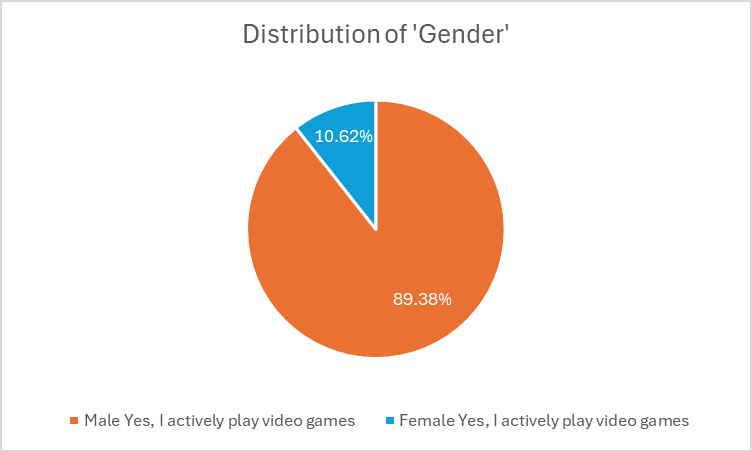}
	\includegraphics[width=0.48\textwidth]{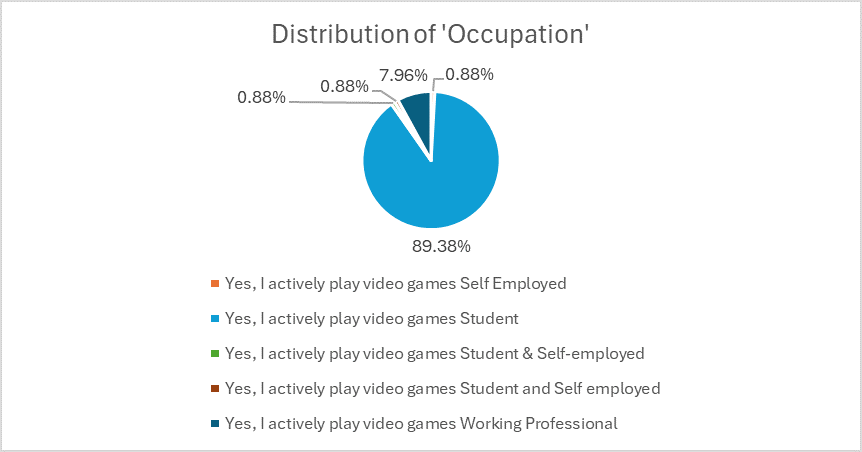}
	\caption{(Above) Gender distribution; (Below) Age distribution of the sample}
	\label{fig:demographics}
\end{figure}
\begin{figure}[h]
	\centering
	\includegraphics[width=0.48\textwidth]{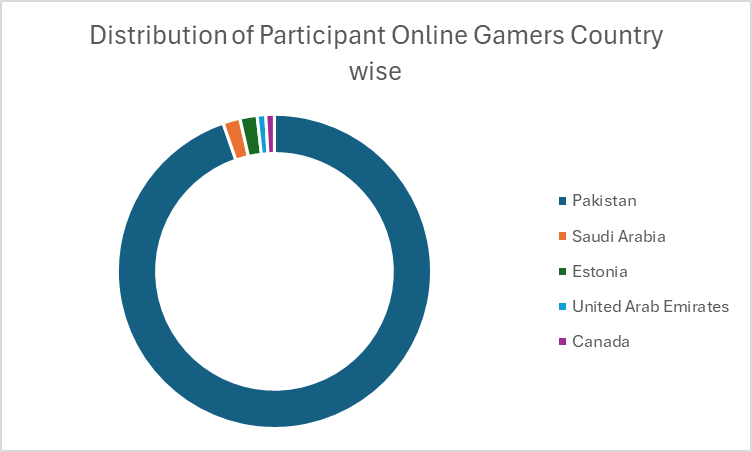}
	\includegraphics[width=0.48\textwidth]{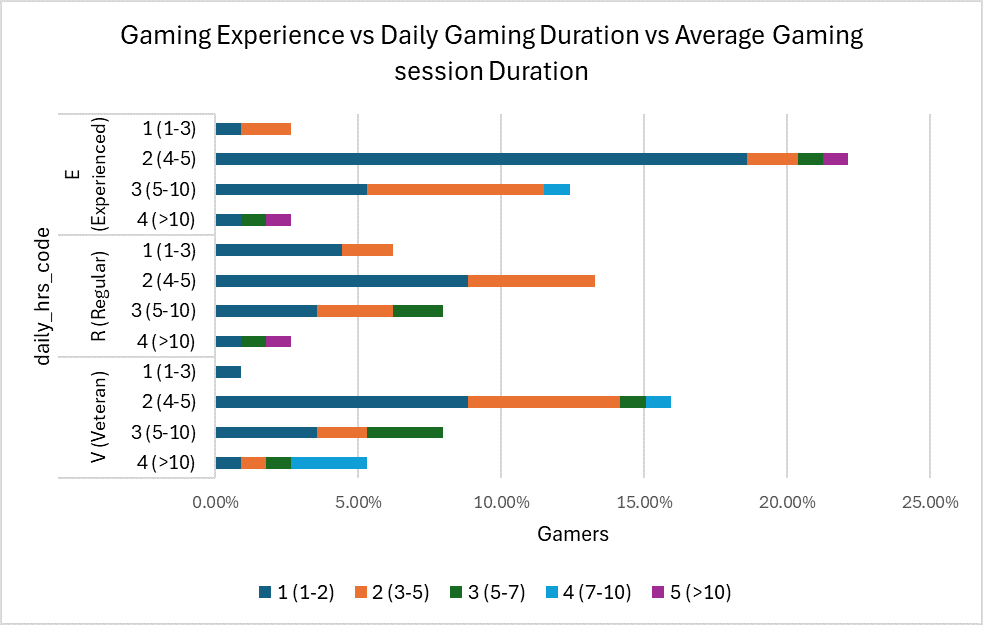}
	\caption{(Above) Participant demographic ; (Below) Gaming time duration }
	\label{fig:demographics}
\end{figure}
\subsection{Correlation Structure and Behavioral Patterns}

Using multiple statistical measures, Spearman's $\rho$ , Cramer's V, Tschuprow's T, Theil’s U, the analysis reveals two main psychological constellations. A salience/rumination axis encompassing escapism, game importance, belonging, and online communication preference. And, an affect/interruptibility axis marked by anxiety on game interruptions and negative mood profiles across social and self-regulation variables.

Behavioral ties show engagement intensity increasing with social activity and motivation for play, with progress tracking central to a high-engagement profile linked to escapism, content creation, and competition. Coherent preference bundles emerge, including a style/achievement-optimization cluster and a skill/learning cluster indicating transfer of life skills through gameplay and community engagement.

Social participation correlates with higher activity and lower negative mood, while content creators demonstrate community engagement and extend participation beyond gameplay. Self-regulation behaviors, such as taking breaks and scheduling, co-occur with positive life experiences and balanced play.

Wellness outcomes are more closely associated with the integration of gaming into daily life routines rather than total playtime; negative effects correlate with escapism and difficulty disengaging, whereas positive outcomes align with structured scheduling and focused skill development, as demonstrated by the strong feature correlations in Figures~\ref{fig:heatmaps} and~\ref{fig:knowledgegraph} and summarized in Tables ~\ref{tab:assoc_partA}, and ~\ref{tab:assoc_partB}.

\begin{table}[htbp]
	\centering
	\caption{Representative Associations Between Psychological, Behavioral, and Social Engagement Variables}
	\label{tab:assoc_partA}
	\small
	\renewcommand{\arraystretch}{1.2}
	\begin{tabular}{|p{4.2cm}|c|c|c|}
		\hline
		\textbf{Variables} & \textbf{Metric} & \textbf{Value} & \textbf{Direction} \\ 
		\hline
		think\_activities $\leftrightarrow$ escapism / belonging / game\_importance & $\rho$ & $\approx +0.62$--$+0.65$ & +ve \\ 
		\hline
		think\_activities $\leftrightarrow$ goal\_setting & $\rho$ & $\approx -0.68$ & -ve \\ 
		\hline
		hard\_disconnecting $\leftrightarrow$ escapism / console / coaching & T & $\approx 0.65$--$0.69$ & — \\ 
		\hline
		mood\_deprivehobby $\leftrightarrow$ goal\_setting / escapism / belonging & $\rho$ & $+0.64 / -0.60 / -0.56$ & Mixed \\ 
		\hline
		irritated\_anxious\_interrupt $\leftrightarrow$ escapism & T; V & $T\approx0.70$; $V\approx0.44$ & — \\ 
		\hline
		neg\_mood\_during $\leftrightarrow$ escapism / community / self\_analysis / goal\_setting & $\rho$ & $-0.62 / -0.50 / +0.48 / +0.46$ & Mixed \\ 
		\hline
		play\_duration $\leftrightarrow$ short\_term\_goal\_salience & $\rho$ & $\approx -0.316$ & -ve \\ 
		\hline
		platform: mobile $\leftrightarrow$ console & V / T / $\rho$ / U & $V=0.389$; $T=0.759$; $\rho\approx-0.18$; $U\approx0.18$--$0.21$ & — \\ 
		\hline
		social\_online $\leftrightarrow$ daily\_activity & $\rho$ & $\approx +0.30$ & +ve \\ 
		\hline
		educational\_score $\leftrightarrow$ tactics / strategy & T & $\approx 0.50$ & — \\ 
		\hline
		like\_custom\_characters $\leftrightarrow$ escapism / coaching / enjoyment / console / self\_analysis & T & $\approx 0.64$--$0.69$ & — \\ 
		\hline
		teach\_lifeskills $\leftrightarrow$ escapism / tactics / tutorial / console / media\_engage & T & $\approx 0.64$--$0.68$ & — \\ 
		\hline
		participate\_ingame\_activities $\leftrightarrow$ escapism / belonging / inspiration / importance / virtual\_first & $\rho$ & $\approx +0.45$--$+0.50$ & +ve \\ 
		\hline
		participate\_ingame\_activities $\leftrightarrow$ goal\_setting / neg\_mood & $\rho$ & $-0.56 / -0.38$ & -ve \\ 
		\hline
		social\_online $\leftrightarrow$ negative\_mood\_during & $\rho$ & $\approx -0.40$ & -ve \\ 
		\hline
		content\_creation $\leftrightarrow$ community\_outside\_game & $\rho$ & $\approx +0.318$--$+0.313$ & +ve \\ 
		\hline
	\end{tabular}
\end{table}

\begin{table}[htbp]
	\centering
	\caption{Representative Associations Between Preferences, Wellness, and Self-Regulation Variables}
	\label{tab:assoc_partB}
	\small
	\renewcommand{\arraystretch}{1.2}
	\begin{tabular}{|p{4.2cm}|c|c|c|}
		\hline
		\textbf{Variables} & \textbf{Metric} & \textbf{Value} & \textbf{Direction} \\ 
		\hline
		take\_gamingbreak $\leftrightarrow$ escapism / pos\_effect\_life / console \& mobile / practice\_competition & T & $\approx 0.60$--$0.64$ & — \\ 
		\hline
		track\_gameprogress $\leftrightarrow$ escapism / content\_creation / competition & T & $\approx 0.71$--$0.76$ & — \\ 
		\hline
		spend\_ingame $\leftrightarrow$ practice / escapism / importance / competition / scheduling & $\rho$ & $\approx +0.39$--$+0.52$ & +ve \\ 
		\hline
		spend\_ingame $\leftrightarrow$ mobile & $\rho$ & $\approx -0.394$ & -ve \\ 
		\hline
		neg\_effect\_life $\leftrightarrow$ hard\_disconnecting / escapism / self\_analysis & $\rho$ & $\approx +0.34$--$+0.62$ & +ve \\ 
		\hline
		pos\_effect\_life $\leftrightarrow$ scheduling / practice / PC-console play & T / $\rho$ & $T\approx0.71$; $\rho$ up to $+0.61$ & +ve \\ 
		\hline
		maintain\_physicalhealth $\leftrightarrow$ bal\_freetime\_offlineactivity & — & — & +ve \\ 
		\hline
		bal\_freetime\_offlineactivity $\leftrightarrow$ escapism / structured practice & T & $\approx 0.62$--$0.65$ & +ve \\ 
		\hline
		scheduled\_activity $\leftrightarrow$ escapism / community / PC-console & T & $\approx 0.62$--$0.76$ & +ve \\ 
		\hline
		priority\_hobby\_responsibility $\leftrightarrow$ solo/story pull / content consumption & — & — & +ve \\ 
		\hline
		game\_imp\_life $\leftrightarrow$ escapism / salience / online-first communication & T & $\approx 0.62$--$0.76$ & +ve \\ 
		\hline
		sleep\_othereffects\_duetohobby $\leftrightarrow$ escapism / track\_progress & T & $\approx 0.66$--$0.69$ & — \\ 
		\hline
	\end{tabular}
\end{table}

%\noindent
\textit{Note:} $\rho$ = Spearman's rank correlation; T = Tschuprow’s T; V = Cramér's V; U = Theil's U. “Positive(+ve)” indicates a direct relationship, “Negative(-ve)” inverse, “Mixed” differing direction across subdimensions.

\begin{figure}
	\centering
	\includegraphics[width=0.88\linewidth]{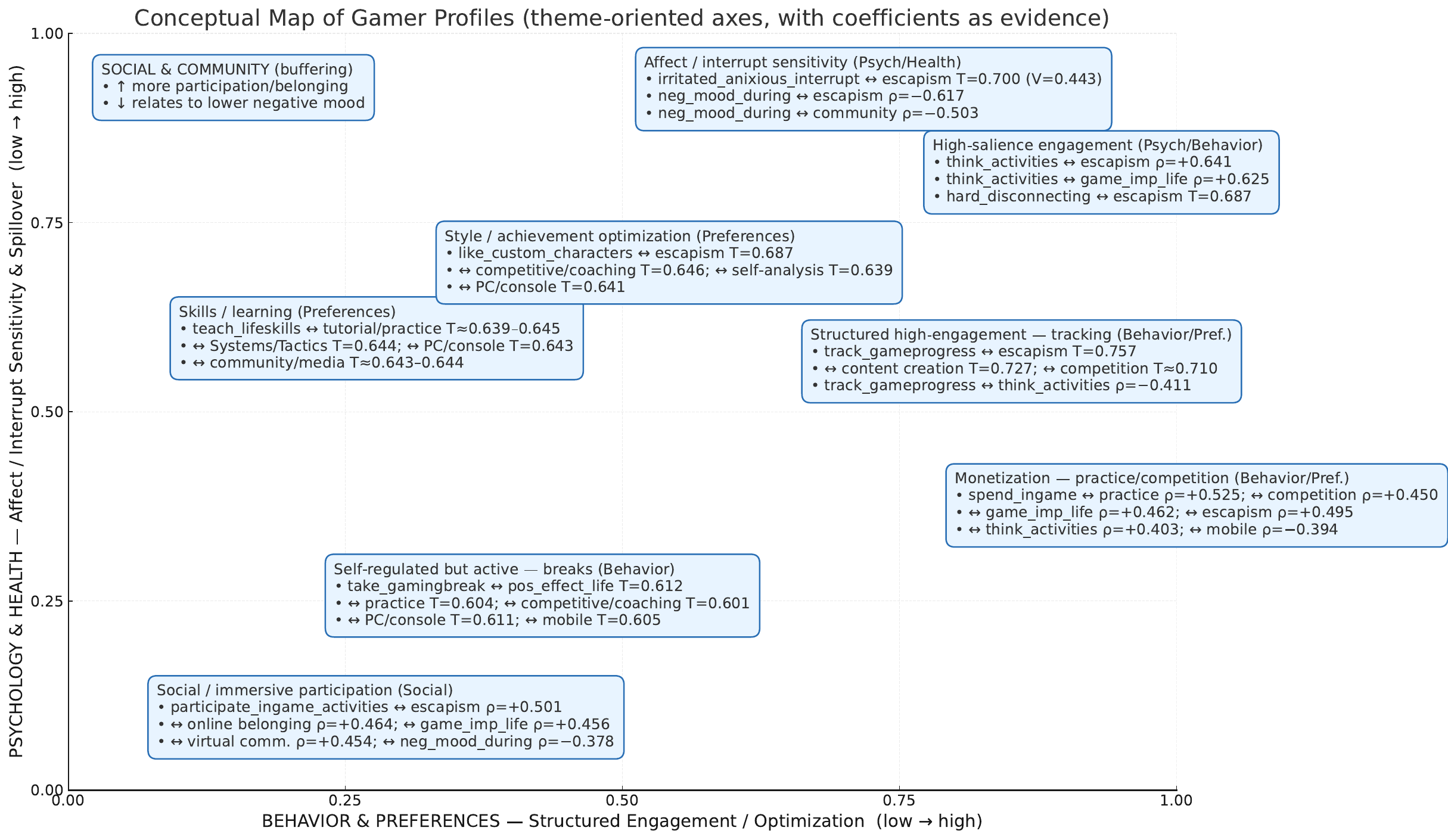}
	\caption{Conceptual map highlighting major dimensions and clusters in gamer profiling.}
	\label{fig:conceptual_map}
\end{figure}

Figure~\ref{fig:conceptual_map} illustrates a conceptual map of gamer profiles along two latent dimensions: \textit{Behavior and Preferences} (structured engagement and optimization) and \textit{Psychology and Health} (affect sensitivity and spillover). Clusters are positioned based on robust associations measured by Cramér’s~V, Tschuprow’s~T, and Spearman’s~$\rho$, reflecting empirically supported relationships among key features.

\subsection{Knowledge Graph Analysis}

The knowledge graph presents as a single, densely connected cluster due to strong correlations among features, resulting in a network where all variables seem globally interconnected. 
However, it consolidates variables into four community clusters that align with psychological, behavioral, preference, social, and wellness axes:

\begin{itemize}
	\item \textbf{Self-Regulation/Health:} Storyline, goal-setting, structured practice, balanced free-time.
	\item \textbf{Skill/Mastery:} Systems/tactics genres, practice, technical hobbies, self-improvement.
	\item \textbf{Story/Content:} Inspiration, mood regulation, physical activity mix, educational cues.
	\item \textbf{Social/Community:} Online socializing, content creation, coaching, progress tracking.
\end{itemize}

This validates thematic clustering and provides rationale for observed personas (Figure~\ref{fig:knowledgegraph}). 

\subsection{Clustering Results and Persona Development}

\begin{table}[htbp]
	\centering
	\caption{Comparative Clustering Metrics for Various $k$ Values}
	\label{tab:clustering_perf}
	\small
	\renewcommand{\arraystretch}{1.2}
	\begin{tabular}{|l|c|c|c|c|}
		\hline
		\textbf{Clustering Technique} & \textbf{k} & \textbf{Silhouette} & \textbf{CH} & \textbf{DB} \\ 
		\hline
		PCA(2D) + K-Means & 3 & 0.413661 & 102.751431 & 0.835544 \\ 
		& 4 & 0.413461 & 106.771112 & 0.797568 \\ 
		\hline
		SVD + K-Means & 3 & 0.077929 & 9.890914 & 2.996473 \\ 
		& 4 & 0.064233 & 7.751610 & 3.235767 \\ 
		\hline
		SVD + Agglomerative & 3 & 0.040821 & 6.638215 & 3.040134 \\ 
		& 4 & 0.039287 & 6.148836 & 2.927230 \\ 
		\hline
		SVD + Spectral & 3 & 0.065781 & 8.659381 & 3.151278 \\ 
		& 4 & 0.055650 & 7.066038 & 3.214607 \\ 
		\hline
		t-SNE(2D) + K-Means & 3 & 0.411735 & 105.751984 & 0.821259 \\ 
		& 4 & 0.403527 & 110.206573 & 0.810651 \\ 
		\hline
	\end{tabular}
\end{table}

\begin{figure}[htbp]
	\centering
	\includegraphics[width=0.7\linewidth]{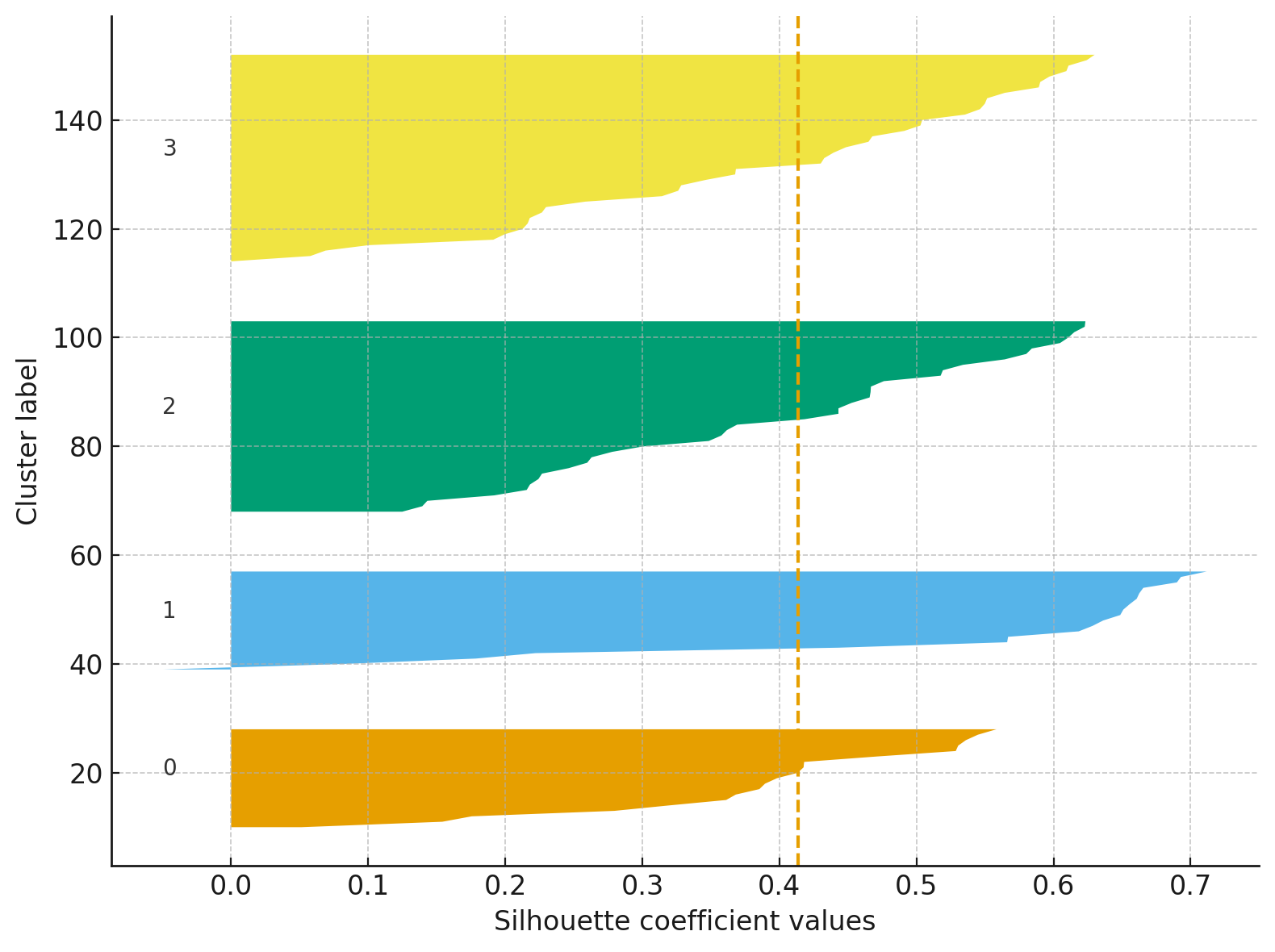}
	\caption{PCA(2D)+K‑Means (k = 4) clusters; four separable personas in 2D projection.}
	\label{fig:PCA+K}
\end{figure}

\begin{figure}[htbp]
	\centering
	\includegraphics[width=0.6\linewidth]{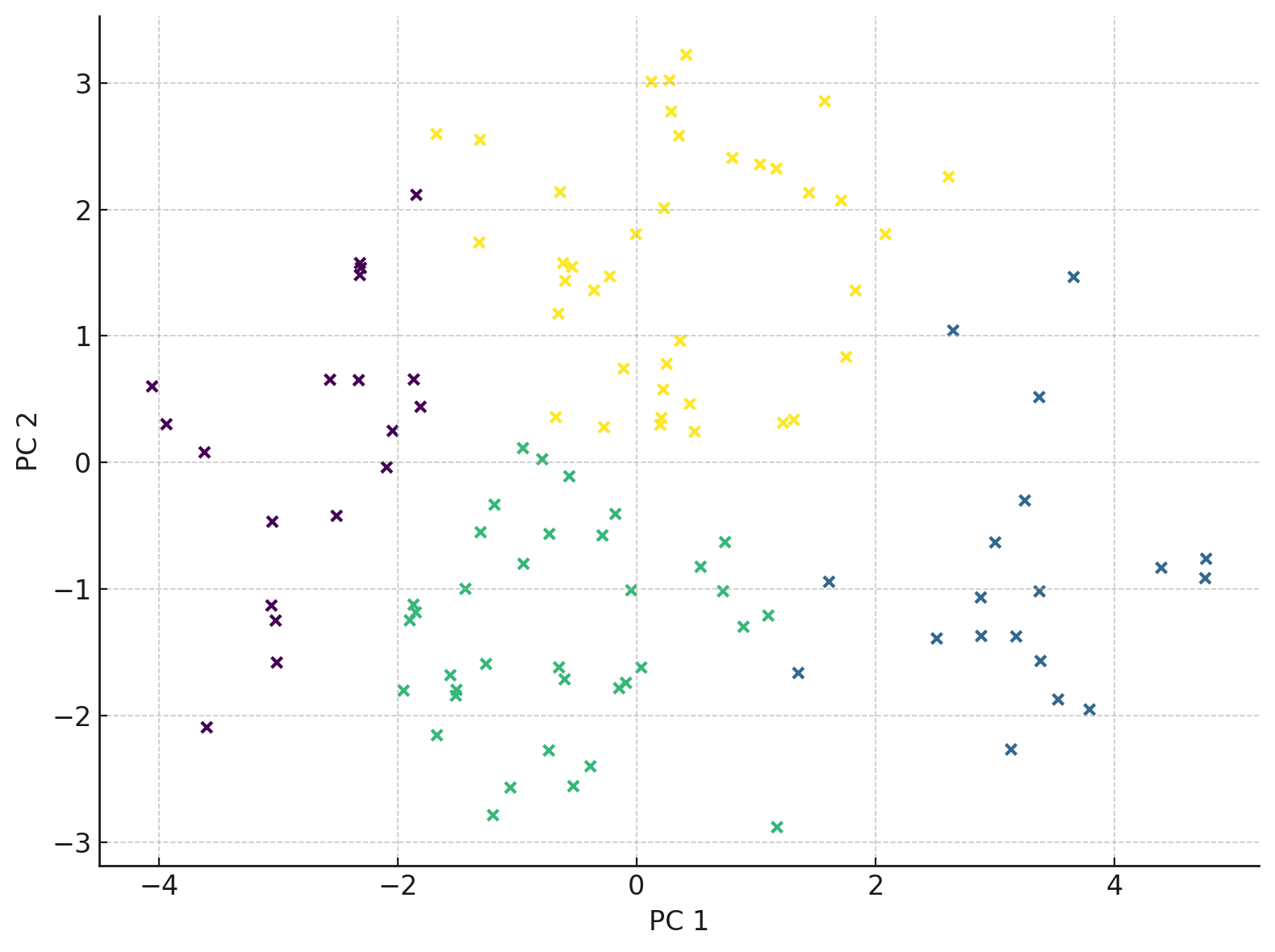}
	\caption{PCA(2D)+K‑Means (k = 4) clusters scatter plot}
	\label{fig:PCA+K2}
\end{figure}

Across methods, $k=4$ consistently balanced separation and compactness. PCA(2D) + K-Means ($k=4$) achieved the lowest Davies--Bouldin (best compactness), silhouette $\approx 0.41$ and CH $\approx 107$, and is thus recommended as the default pipeline as in Tab.~\ref{tab:clustering_perf}.

t‑SNE(2D)+K‑Means produced similar Silhouette but a slightly worse DB; SVD‑based pipelines exhibited weak separation.  Fig.~\ref{fig:PCA+K} and ~\ref{fig:PCA+K2} shows the silhouette plot for the KMean based clusters and scatter plot for PCA 1 and 2 . Cluster 1 shows slight overlapping and Cluster 2 is more dominant cluster.  Each color represents a unique cluster characterized by shared behavioral and psychological traits. The clear separation and compactness of clusters indicate the efficacy and robustness of the unsupervised segmentation approach, serving as a foundation for targeted interventions and personalized game design. 

Stability checks indicated lower adjusted‑Rand at small N but acceptable within‑cluster Jaccard. A simple downstream logistic Regression Classifier  achieved ~90\% accuracy when predicting cluster membership from features, suggesting that broad tendencies are learnable despite local assignment variability.

\subsection{Gamer Persona Profiles}

Interpreting the four segments against the correlation communities yields coherent gamer personas. These are not strict causal categories but reproducible constellations that reflect structured engagement, social immersion, affective or interruption sensitivity, and style- or achievement-oriented pathways. Table~\ref{tab:personas} summarizes these personas along with their defining features and potential design implications.

\begin{table}[h]
	\centering
	\caption{Gamer Personas: Definitions and Engagement Considerations}
	\begin{tabular}{|p{2cm}|p{0.8cm}|p{2cm}|p{2cm}|}
		\hline
		Persona & Cluster & Design Levers & Notes \\
		\hline
		Social Explorers & C0 & Community events, co-op play, creator tools & Lower negative mood during play; connection framing\\
		\hline
		Discipline Optimizer & C1 & Telemetry support, coach based learning, habit scaffolds & Slightly higher arousal; slightly lower valence \\
		\hline
		Strategic Navigator & C2 & Mastery challenges, narratives & Bridges skill and story orientations\\
		\hline
		Competitive Socializer & C3 & Ranked ladders, mentorship & Narrower repertoires; high optimization drive \\
		\hline
	\end{tabular}
	\label{tab:personas}
\end{table}

\section{Comparative Analysis of Gamer Profiles}

This research identifies the four distinct gamer segments; Cluster-0 Social Explorers, Cluster-1 Discipline Optimizer, Cluster-2 Strategic Navigator, and Cluster-3 Competitive Socializers; and aligns them with established motivational and typological frameworks emphasizing sociality, mastery, immersion, metagame engagement, and wellness.

Our C3 cluster corresponds to competition and achievement motives within MOGQ %\cite{demetrovics2011you}
and Yee’s models %\cite{yee2006motivations}; 
,C0 reflects social and immersion drives; C2 integrates mastery with fantasy and immersion, %\cite{nacke2014brainhex, veracruz2023bigfive};
and C1 highlights optimization and self-regulation. %\cite{kiraly2022comprehensive}
These align well with classic typologies such as Bartle’s Achiever/Conqueror, %\cite{bartle1996mud}
Socializer/Explorer, and BrainHex’s Mastermind/Seeker, %\cite{nacke2014brainhex}
with C1 adding explicit behavioral regulation signals.

Contemporary research reinforces these mappings: Vera Cruz
%~\cite{veracruz2023bigfive} 
identify Mastermind and Seeker as dominant archetypes matching our systems/narrative segment, the Gaming Motivation Inventory %\cite{kiraly2022comprehensive}
distinguishes adaptive and maladaptive motives complementing our wellness clusters, metagamer profiles emphasize strategic play and coaching resonant in C3 
%\cite{kahila2023metagamers},
, and emotional network analyses support C0's intrinsic motivation-driven engagement %\cite{george2024emotional}
. The results are summarized in the Table~\ref{tab:comparision}.	
\begin{table}[htbp]
	\centering
	\caption{Points of Convergence and Extension Relative to Selected Studies}
	\label{tab:comparision}
	\small
	\renewcommand{\arraystretch}{1.2}
	\begin{tabular}{|p{1.3cm}|p{2cm}|p{2cm}|p{1.8cm}|}
		\hline
		\textbf{Feature Domain} & \textbf{This Study — Segment Signal} & \textbf{Comparable External Findings} & \textbf{What This Study Adds} \\
		\hline
		Systems / Mastery &
		systems/tactics + narrative; structured practice &
		BrainHex Mastermind; MOGQ/GMI mastery factors &
		Combines systems with story + wellness indicators \\
		\hline
		Competition / Social &
		competitive practice, coaching, progress tracking &
		Yee Achievement+Social; GMI competition; Metagamers Strategizers &
		Observed metagame behaviors (coaching/tools), not just motives \\
		\hline
		Immersion / Narrative &
		social + narrative engagement; lower negative mood &
		Yee Immersion; BrainHex Seeker; emotional-attachment networks &
		Links immersion to mood buffering and belonging \\
		\hline
		Regulated Engagement &
		scheduling, breaks, tracking; positive wellness &
		GMI adaptive vs. maladaptive motives; SDT competence/autonomy &
		Actionable self-regulation signals with wellness outcomes \\
		\hline
		Wellness Outcomes &
		Positive vs. negative life effects mapped to motives/behaviors &
		GMI and SDT/PENS associations with well-being &
		Direct wellness features tied to personas; design implications \\
		\hline
	\end{tabular}
\end{table}

This synthesis integrates classical and emerging perspectives, validating our data-driven profiles and highlighting novel behavioral markers.
\section{Study Contributions and Future Directions}

This study demonstrates that robust segmentation of online gamers is achieved by integrating behavioral telemetry, psychological self-reports, preference profiles, and social interaction data, all grounded in validated motivational frameworks such as the MOGQ and SDT. Unsupervised machine learning on this multimodal feature set produced interpretable personas mapped to axes of self-regulation, mastery, narrative immersion, and competitive socialization.

The analysis answers RQ1 and RQ2 by revealing the gamer types, distinguished by their engagement structure, motivational orientations, affective sensitivity, and social as well as content creation patterns. Profiles range from highly organized, progress-tracking optimizers to socially adaptive participants, interruption-sensitive players, and style-oriented achievers. These segments show different associations with well-being, with adaptive engagement and self-regulation supporting positive outcomes and ruminative routines carrying risk.

Machine learning techniques using K-mean clustering principal component analysis, knowledge graph community detection, and K-means clustering uncovered stable latent structures linking engagement, self-regulation, and well-being. Predictive modeling with logistic regression confirmed the stability and discriminability of emergent personas, with accuracy rates near 90\%. Importantly, integration of behavioral markers such as scheduling, regular breaks, and progress tracking helped distinguish adaptive from maladaptive play, while the knowledge graph elucidated meso-level communities explaining cluster emergence, thus this answer RQ3.

This work bridges classical gamer typologies with direct behavioral markers, psychological, social and wellness indicators, illuminating design levers for health- and balance-oriented play. Practical implications include targeting interventions, enabling personalized feedback, and guiding future game experiences. 

Although supervised prediction of segment membership indicates good generalization, internal validity indices can be sensitive to class imbalance or the scaling of mixed-type variables, which may be a consequence of the limited data.

Directions for further research include extending analysis to cross-cultural settings, performing longitudinal tracking of player evolution, and enriching segmentation with enhanced multimodal and temporal data streams.
\section*{Conflict of Interest}
The authors declare that there is no conflict of interest.

\section*{Data Availability}
The data that support the findings of this study are available from the corresponding author upon reasonable request.

\section*{Acknowledgment}
The authors would like to acknowledge the valuable assistance of Syed Ali Hasnat and Michael Ghori, undergraduate students at Iqra University, Khi-PK, for their contributions in data collection and data cleaning. The authors also acknowledge the use of AI-assisted tools, including ChatGPT and similar platforms, which were utilized to support literature review exploration and manuscript preparation.

\section*{Acknowledgement(s)}

	The authors would like to acknowledge the valuable assistance of Syed Ali Hasnat and Michael Ghori, undergraduate students at Iqra University, Khi-PK, for their contributions in data collection and data cleaning. The authors also acknowledge the use of AI-assisted tools, including ChatGPT and similar platforms, which were utilized to support literature review exploration and manuscript preparation.

\section*{Disclosure statement}

The authors declare that there is no conflict of interest.

\section*{Funding}

There is no funding related research

\section*{Notes on contributor(s)}

{\textbf{MOONA KANWAL}}  ({https://orcid.org/0000-0001-7656-373X}) received her PhD (Computer Science) degree in 2024 from NED University of Engineering and Technology, Pakistan. She completed her BSc. in 2002 and an MSc in computer engineering in 2007 from Sir Syed University of Engineering and Technology Pakistan. She is currently serving Iqra University Pakistan as HoD Software Engineering Dept. She has worked as Assistant Professor at Sir Syed University of Engineering and Technology Pakistan since 2007. She is also a mentor for many projects and startups. She has also held the position of Professor (part-load) at Seneca College Canada from 2020 to 2022. She has also worked as Compliance and SQA Manager at Mixit Technologies Pakistan from 2008 to 2009. Ms. Moona received funding from the Ministry of Science and Technology Pakistan and a scholarship from the Higher Education Commission Pakistan. Her research interests include AI, data sciences, human-computer interaction, and Human behavior.
\\
{\textbf{MUHAMMAD SAMI SIDDIQUI}}
({https://orcid.org/0000-0002-4937-6266}{ORCID: 0000-0002-4937-6266}) received the Ph.D. degree in Mathematics from the Institute of Business Administration (IBA), Karachi, Pakistan, in 2023. He is currently serving as an Assistant Professor at Iqra University, Pakistan. His research interests include nonlinear dynamics, bifurcation theory, and computational mathematics, with recent work focusing on the integration of data-driven and machine learning techniques. At present, Dr. Siddiqui is involved in projects on gamer profiling and fraud detection using machine learning methods. He is also part of an international collaborative research group working on formalizing mathematical theorems in Lean4's MathLib. He received a research travel grant from IBA Karachi under SIAM to attend SWAM 2022 in Charles University, Prague, where he presented his research work. He has also presented his work at international academic events hosted in Ukraine, Georgia, and Bulgaria.

{\textbf{SYED ANAEL ALI}}
an A-level student specializing in Physics and Mathematics, is currently pursuing studies in Computer Science from Karachi Grammer School, PK. His research interests encompass video game development and data analytics. He recently contributed to initial research on video gamer profiling and was responsible for writing the introduction and literature review sections, with a particular focus on these areas.

	\section*{Data Availability}
The data that support the findings of this study are available from the corresponding author upon reasonable request.

\bibliographystyle{tfq}
\bibliography{references}

\begin{thebibliography}{10}
\bibitem{pearson1900x2}
K. Pearson, \emph{On the criterion that a given system of deviations from the
  probable in the case of a correlated system of variables is such that it can
  be reasonably supposed to have arisen from random sampling}, Vol.~50, 1900.
\end{thebibliography}

\end{document}